\documentclass[aps,prx,twocolumn,superscriptaddress,bibnotes]{revtex4-1} 

\usepackage{amsmath}
\usepackage{amssymb}
\usepackage{amsthm}
\usepackage{bbold,bm}
\usepackage[cal=boondox]{mathalfa}
\usepackage{graphicx}

\newcommand\vex[1]{\mathbf{#1}}
\newcommand\gvex[1]{\boldsymbol{#1}}

\def\ket#1{\mathinner{|{#1}\rangle}}

\def\braket#1{\mathinner{\langle{#1}\rangle}}

\def\re{\mathrm{Re}\,}
\def\im{\mathrm{Im}\,}

\def\id{\mathbb{1}} 

\def\Texp{\mathrm{T}\hspace{-0.5mm}\exp}

\def\nodag{{\vphantom{\dagger}}}

\usepackage{color}
\def\noteI#1{{#1}}
\def\noteII#1{{#1}}

\newcommand{\extra}[1]{\textcolor{cyan}{}}

\begin{document} 

\title{Higher-order Floquet topological phases with corner and bulk bound states} 

\author{Martin Rodriguez-Vega}
\affiliation{Department of Physics, Indiana University, 727 E Third Street, Bloomington, Indiana 47405, USA}
\affiliation{Department of Physics, The University of Texas at Austin, Austin, TX 78712, USA}
\affiliation{Department of Physics, Massachussetts Institute of Technology, Cambridge, MA 02139, USA}

\author{Abhishek Kumar}
\affiliation{Department of Physics, Indiana University, 727 E Third Street, Bloomington, Indiana 47405, USA}

\author{Babak Seradjeh}
\email[Corresponding author, ]{babaks@indiana.edu.}
\affiliation{Department of Physics, Indiana University, 727 E Third Street, Bloomington, Indiana 47405, USA}
\affiliation{Max Planck Institute for the Physics of Complex Systems, N\"othnitzer Stra\ss e 38, Dresden 01187, Germany}

\begin{abstract}
We report the theoretical discovery and characterization of higher-order Floquet topological phases dynamically generated in a periodically driven system with mirror symmetries. We demonstrate numerically and analytically that these phases support lower-dimensional Floquet bound states, such as corner Floquet bound states at the intersection of edges of a two-dimensional system, protected by the nonequilibrium higher-order topology induced by the periodic drive. We characterize higher-order Floquet topologies of the bulk Floquet Hamiltonian using mirror-graded Floquet topological invariants. This allows for the characterization of a new class of higher-order ``anomalous'' Floquet topological phase, where the corners of the open system host Floquet bound states with the same as well as with double the period of the drive. Moreover, we show that bulk vortex structures can be dynamically generated by a drive that is spatially inhomogeneous. We show these bulk vortices can host multiple Floquet bound states. This ``stirring drive protocol'' leverages a connection between higher-order topologies and previously studied fractionally charged, bulk topological defects. Our work establishes Floquet engineering of higher-order topological phases and bulk defects beyond equilibrium classification and offers a versatile tool for dynamical generation and control of topologically protected Floquet corner and bulk bound states.
\end{abstract}

\maketitle 

\section{Introduction}
Topological phases of matter are characterized by an intimate relationship between the patterns of motion in the bulk and those at the boundaries of the system. While there is no general theory of this bulk-boundary correspondence, it is known to hold in certain classes of topological phases, e.g. those of non-interacting fermions protected by internal and crystalline symmetries~\cite{SchRyuFur08a,kitaev2009,RyuSchFur10a,TeoKan10a,fu2011}. The interface between two such phases with the same symmetries and different topological invariants binds gapless modes. For example, a one-dimensional interface between the two-dimensional quantum spin Hall phase and a trivial insulator supports an odd number of gapless helical edge modes~\cite{KanMel05a,BerHugZha06a,HasKan10a}.

In recent years, topological classification of phases of matter has been extended to systems that are driven periodically out of equilibrium~\cite{Fru16a,roy2016,yao2017}. In these systems, bulk-boundary correspondence acquires a new, temporal character: for a drive frequency $\Omega$, the localized boundary modes may now coexist at the same interface at different values of the quasienergy, $\epsilon^+=0$ (Floquet zone center) and $\epsilon^-=\Omega/2$ (Floquet zone edge)~\cite{RudLinBer13a,NatRud15a}. At high frequencies, the dynamics self-averages to equilibrium and the quasienergies asymptotically approach the energies of the average Hamiltonian. Thus, the boundary modes at the quasienergy zone edge disappear and the Floquet topology coincides with the equilibrium topology of the average Hamiltonian. As the frequency is lowered, topological transitions at the Floquet zone edge and center are induced and Floquet topology acquires a richer structure than any equilibrium topology~\cite{OkaAok09a,jiang2011,LinRefGal11a,GuFerAro11a,kundu2013,tong2013,KunFerSer14a,vega2018}. Apart from their richer topological structure, Floquet topological phases promise practical advantages over their equilibrium counterparts, such as more control. Indeed, the topological phase of the system can be tuned, usually with great precision, by the \emph{drive protocol} (drive amplitude, frequency, and shape), thus allowing phase transitions in situ. 

More recently, the notion of bulk-boundary correspondence has been generalized to higher-order topological phases in equilibrium, whose surfaces at one lower dimension remain gapped, yet support gapless modes localized at their lower-dimensional boundaries, such as hinges and corners~\cite{benalcazar2017a,benalcazar2017b,SchCooVer18a,langbehn2017,son2017}. For example, a two-dimensional electric quadrupole topological insulator binds corner states with fractional charge $e/2$. Such higher-order topological phases have been predicted to exist in engineered lattices of cold atoms~\cite{benalcazar2017b} and in natural elemental bismuth~\cite{schindler2018b}, and have been observed in a mechanical system of coupled microwave resonators~\cite{PetBenHug18a,SerPerSus18a}, optical waveguides~\cite{NohBenHua18a}, topolectrical circuits~\cite{imhof2018}, mechanical metamaterials~\cite{serra2018}, and elastic acoustic structures~\cite{FanXiaTon18a}. In this work, we show that \emph{higher-order Floquet topological phases} can be realized and controlled in a periodically driven system, supporting lower-dimensional Floquet bound states at the Floquet zone center and/or edge. 

Specifically, we study a driven model with mirror symmetries that realizes Floquet topological quadrupole phases and supports Floquet corner states. We show that with open boundary conditions this system supports Floquet bound states at the corners. With periodic boundary conditions, 
we characterize these phases using mirror-graded Floquet topological invariants. In particular, we show these invariants correctly predict the higher-order \emph{anomalous} Floquet topological phase that supports Floquet corner states at both Floquet zone center \emph{and} edge.

Furthermore, we study drive protocols that are spatially inhomogeneous. We design specific protocols that can be used to ``stir'' topological bulk defects, namely vortices, that host lower-dimensional Floquet bound states in the bulk. Thus, we expand the notion of higher-order topology to systems with spatiotemporal nonuniformities. 

The paper is organized as follows. In Section~\ref{sec:model}, we review the model, its symmetries, and the characterization of its equilibrium higher-order topology. Here, we also introduce our notation of Floquet theory and the general scheme of defining Floquet topological invariant for a drive protocol with time-reflection symmetry. In Section~\ref{sec:HOFT}, we use this scheme to study the driven model and characterize, analytically and numerically, the higher-order Floquet topology as a function of frequency. In Section~\ref{sec:stir}, we introduce spatially inhomogeneous drive protocols that stabilize bulk vortex structure supporting Floquet bound states localized at their cores. We conclude in Section~\ref{sec:discussion} with a discussion and outlook for future work. \noteI{We present some details of our calculation and arguments in two Appendices.}

\vspace{-2mm}
\section{Model and Floquet Theory}\label{sec:model}
In this section, we introduce the model that exhibits higher-order topological phases with an emphasis on the algebra of its symmetries. For completeness, we also briefly review the method of characterizing its higher-order topology in equilibrium in the presence of certain symmetries as well as the Floquet theory of periodic dynamics. This will set the stage for describing higher-order Floquet topologies in Section~\ref{sec:HOFT}.

\vspace{-3mm}

\subsection{Model}
\begin{figure}[t]
   \centering
   \includegraphics[width=3.4in]{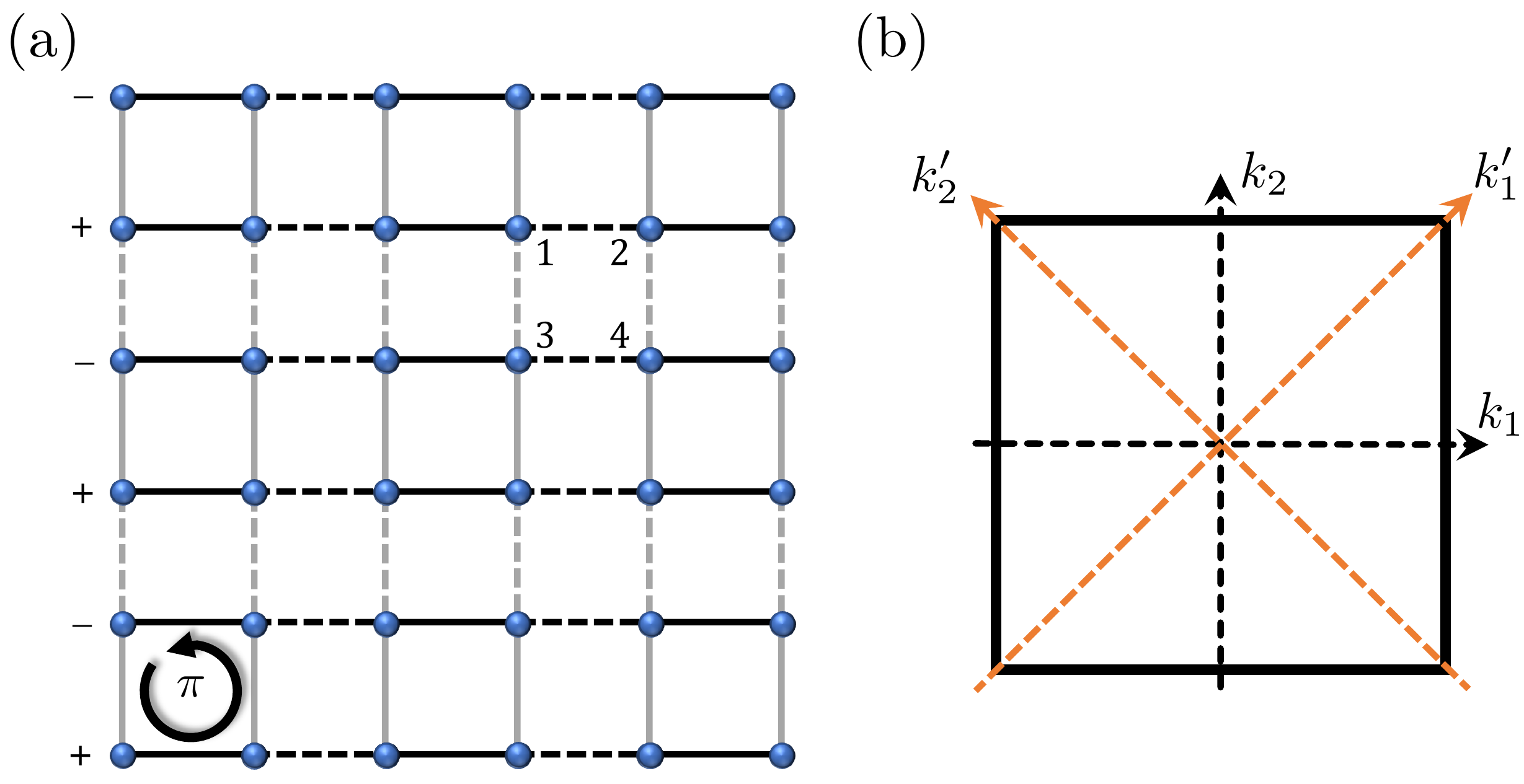} 
   \caption{\noteI{(a) The $\pi$-flux dimerized square lattice with spatially modulated hopping amplitudes. The elements within a unit cell are labeled 1 through 4, and each plaquette carries a $\pi$ flux, which may be represented in the Landau gauge by an alternating pattern of $\pm$ for hopping amplitudes along the rows.
   (b) The Brillouin zone showing the principal ($k_1$ and $k_2$) and the diagonal ($k'_1$ and $k'_2$) mirror symmetry directions.}}
   \label{fig:HOFTsLattice}
\end{figure}

We demonstrate our findings in a driven $\pi$-flux dimerized square lattice as a minimal model of a two-dimensional quadrupole Floquet topological insulator. The Hamiltonian is~\cite{seradjeh2008}
\begin{equation}
H = \sum_{\braket{\vex r \vex s}} w_{\vex r \vex s}e^{i\phi_{\vex r \vex s}} c^\dagger_\vex r c^\nodag_\vex s 
\end{equation}
where $c_\vex r$ annihilates a spinless fermion at site $\vex r=(x,y)$, 
$w_{\vex r \vex s}=w^*_{\vex s \vex r}$ are hopping amplitudes between nearest neighbors $\braket{\vex r\vex s}$, and $\phi_{\vex r \vex s}$ are Peierls phases implementing the magnetic flux penetrating the lattice. In the Landau gauge $\phi_{\vex r \vex r+\vex e_\mu} = \pi y \hat{\vex x}\cdot \vex e_\mu$, where we have used natural units $\hbar=c=e=1$. The hopping amplitude in the direction $\vex e_\mu$ of a nearest neighbor is modulated as
\begin{equation}\label{eq:hop}
w_{\vex r \vex r+\vex e_\mu} = w_\mu[1 - \re( \eta^*_{\vex r \mu} f_\vex r )],
\end{equation}
where $\eta_{\vex r \mu} = e^{i \arg \vex e_\mu} e^{i \pi \vex r \cdot \vex e_\mu}$ are directional complex signs~\cite{seradjeh2008}. Here, $f_\vex r$ is a complex function that specifies the hopping modulation locally. For a uniform $f_\vex r = f = |f|e^{i\chi}$ we have
\begin{align}
w_{\vex r \vex r \pm \hat{\vex x}} &= w_1[1 \mp (-1)^x |f|\sin\chi ],\\
w_{\vex r \vex r \pm \hat{\vex y}} &= w_2[1 \mp (-1)^y |f|\cos\chi ],
\end{align}
i.e. a uniform hopping modulation by $\re f$ ($\im f$) in the $x$ ($y$) direction. \noteII{See Fig.~\ref{fig:HOFTsLattice} for a depiction of the model and the mirror symmetry axes.}

In this case, the unit cell has four basis points; thus, for a unit cell at position $\vex R$ with the site $\vex r$ at its corner, $\psi_{\vex R}^\dag = (c^\dag_{\vex r},c^\dag_{\vex r+\hat{\vex x}}, c^\dag_{\vex r+\hat{\vex y}}, c^\dag_{\vex r+\hat{\vex x}+\hat{\vex y}})$ defines a unit-cell spinor. 
For a system with $L$ sites and periodic boundary conditions, we can write the Hamiltonian in the Bloch basis $\psi_{\vex k}^\dagger = \frac1Le^{-(\pi/4)\gamma_2}\sum_{\vex R} e^{i\vex k\cdot\vex R}\psi_{\vex R}^\dagger$ with lattice momentum $\vex k = (k_1,k_2)$, as $H = \sum_{\vex k} \psi_{\vex k}^\dagger H(\vex k) \psi^\nodag_{\vex k}$, with Bloch Hamiltonian
\begin{align}\label{eq:piS}
H(\vex k) 
	= A_1 D_1(k_1) + A_2 D_2(k_2).
\end{align}
Here, $A_j = \gamma_0\gamma_j$, $D_j = |d_j|e^{i\phi_j C_j}$ for $j=1,2$, with
\begin{align}
|d_j|e^{i\phi_j}
	&= w_j[1+f_j+(1-f_j)e^{ik_j}] \nonumber \\
	&= 2w_je^{ik_j/2}\left(\cos\frac{k_j}2-if_j\sin\frac{k_j}2 \right),
\label{eq:dj}
\end{align}
where $f_1 \equiv \re f$, $f_2 \equiv \im f$, $C_1 = -i\gamma_1$, $C_2=\gamma_2\gamma_5$, and $\gamma_\alpha$, $\alpha\in\{0,1,2,3\}$, are Dirac matrices satisfying the Clifford algebra $\{\gamma_\alpha , \gamma_\beta \} =2g_{\alpha\beta}$ with the metric $g = \text{diag}(1,-1,-1,-1)$, and $\gamma_5 = -i\gamma_0\gamma_1\gamma_2\gamma_3$. We use the Weyl basis $\gamma_0=\sigma_1\otimes\id$, $\gvex\gamma=i\sigma_2\otimes\gvex\sigma$, and $\gamma_5=\sigma_3\otimes\id$ in terms of Pauli matrices $\gvex\sigma$.

%
We may also write the Hamiltonian as $H = \sum_\alpha \Gamma_\alpha \mathcal{d}_\alpha $, where the anticommuting matrices $\Gamma_0 = A_1$, $\Gamma_1 = iA_1C_1$, $\Gamma_2 = A_2$, $\Gamma_3 = iA_2C_2$, and the 4-vector 
\begin{equation}
\mathcal{d} = (|d_1|\cos\phi_1, |d_1|\sin\phi_1, |d_2|\cos\phi_2, |d_2|\sin\phi_2).
\end{equation}
There are four bands with doubly degenerate energies $\pm E(\vex k)$, where $E(\vex k)=|\mathcal{d}(\vex k)|$.

\vspace{-3mm}
\subsection{Symmetries}
The above model is a two-dimensional generalization of the one-dimensional Su-Schrieffer-Heeger (SSH) model \cite{su1979} equipped with the proper Clifford algebra~\cite{CalJurRoy18a}. Here, we focus on the algebra of the symmetries that determine its spectral and topological properties regardless of the choice of unit-cell basis or gauge. The Hamiltonians
\begin{equation}
H_j(k_j) \equiv A_j D_j(k_j),
\end{equation}
represent a SSH model in the $k_j$ direction. In each direction, the Hermitian unitary operator $C_j = C_j^\dagger=C_j^{-1}$ represents a chiral symmetry:
\begin{equation}
\{A_j, C_j\}=0 \Rightarrow C_j H_j(k_j) C_j = -H_j(k_j).
\end{equation}
The commutation algebra
\begin{equation}
[C_1,C_2]=[C_1, A_2]=[C_2, A_1] = \{ A_1 ,A_2 \} = 0
\end{equation}
ensures that these two SSH Hamiltonians \emph{anticommute} with each other, $\{H_1,H_2\}=0$. 
Consequently, the operator $C = C_1 C_2 = \gamma_0\gamma_3$, $\{ C, A_j \}$, is the chiral symmetry of the full Hamiltonian $H=H_1+H_2$, $\{ C, H\} = 0$. 

Due to the enlarged dimension of the unit cell, each direction now has a continuum of discrete symmetries. For example, mirror symmetries $M_j H_j(k_j) M_j^{-1} = H(-k_j)$ are given by $M_j = A_j U_{j}$, where $U_{j}$ is a unitary that commutes with $H_{j}$. This is the U(2) group generated by $\{ \id, C_{\bar j}, A_{\bar j} C_j, i A_{\bar j} C\}$, where $\bar j \neq j$ is the complement of $j$. Imposing the condition $[M_j, H_{\bar j}] = 0$ then chooses $M_j = A_j C_{\bar j}$ as the mirror symmetry of the full Hamiltonian. Thus, $M_1=i\gamma_3$ and $M_2 = \gamma_3\gamma_5$. We have
\begin{equation}
\{M_1,M_2\} 
= \{ M_j , C_j\} = [M_j, C_{\bar j}] 
= [M_j, A_{j'}]
= 0.
\end{equation}
Since $M_1$ and $M_2$ anticommute, $I=-iM_1M_2=\gamma_5$ is a Hermitian unitary representing the inversion symmetry. 

We note the action of two diagonal mirror symmetries, $M'_1:(k_1,k_2)\mapsto(-k_2,-k_1)$ and $M'_2:(k_1,k_2)\mapsto(k_2,k_1)$. Supplied with proper Hermitian unitaries,
\begin{align}
M'_1 &= e^{i(\pi/4) A_3}e^{i(\pi/4) I} M_1A_2, 
\label{eq:diagM1}
\\
M'_2 &= e^{i(\pi/4) A_3}e^{i(\pi/4) I} C_1,
\label{eq:diagM2}
\end{align}
where $A_3 = -iA_1A_2$,
these operations connect different Bloch Hamiltonians by mapping the hopping modulation vector $\vex f \equiv (f_1,f_2)\mapsto(f_2,f_1) = M'_2\vex f$ and the average hopping vector $\vex w \equiv (w_1,w_2) \mapsto (w_2,w_1) = M_2'\vex w$,
\begin{equation}
M'_j H(\vex k; \vex w, \vex f) {M'_j}^{-1} = H(M'_j \vex k;M'_2\vex w, M'_2\vex f).
\end{equation}
Note that
\begin{equation}
[M'_j,C] = \{ M'_1, M'_2 \} = 0.
\end{equation}
For $w_1=w_2$ and $f_1=f_2$, the model has diagonal symmetries $M'_j$ as well as a four-fold rotational symmetry generated by $R_4=M_2M'_1$.

This model also has antiunitary paticle-hole symmetry, $P H(\vex k) P^{-1} = -H^*(-\vex k)$ with $P = P^{-1} = \gamma_1\gamma_5 K$, and time-reversal symmetry, $T H(\vex k) T^{-1} = H^*(-\vex k)$ with $T = T^{-1} = i CP = \gamma_2 K$, where $K$ is the complex conjugation operator. Therefore, the model belongs to the BDI Altland-Zirnbauer class~\cite{altland1997}. These symmetries satisfy
\begin{align}
&\{P, M_j \} = [T,M_j] = 0,\\
&\{P,T\} = [P,C] = [T,C] =0.
\end{align}

\subsection{Higher-Order Topological Phases}
In equilibrium, when all parameters are time-independent, and for uniform modulation $f=|f|e^{i\chi}$, the model exhibits four phases: a trivial insulating phase for $0<\chi<\pi/2$, a $x$- or $y$-edge-polarized insulating phase for $0<\chi\mp\pi/2<\pi/2$, and a second-order topological insulating phase for $-\pi<\chi<-\pi/2$. One can tune between the trivial and second-order topological phases by closing the bulk gap at $m=0$. However, one may also keep the bulk gap open, $m\neq 0$, and cross between the trivial and edge-polarized, or edge-polarized and topological quadrupole phases. With open boundary conditions, the gap closes in the edge spectrum, and the second-order topological phase show corner bound states. 

With periodic boundary conditions, these phases have been characterized~\cite{benalcazar2017a, benalcazar2017b} in terms of nested Wilson loops of Wannier bands. A simpler characterization of these phases becomes possible when the model admits diagonal mirror symmetries $M'_j$, i.e. when $f_1=f_2$ and $w_1=w_2$. In this case, a bulk mirror-graded topological invariant~\cite{jeffrey2008,fu2011,TopoMatter} can be defined as
\begin{equation}\label{eq:mirrinv}
\nu_{j} = \frac{\nu_{j+} - \nu_{j-}}2,
\end{equation}
where $\nu_{j\pm}$ are the winding numbers of $H_{j\pm}(\vex k'_{mj})$, the Hamiltonian projected on the mirror eigenspace of $M'_j$ with eigenvalues $\pm1$, along the symmetric lines of $M'_j$ in the Brillouin zone, $\vex k'_{m1} = (k,-k)$ and $\vex k'_{m2}=(k,k)$. For an open system, the invariant line under diagonal mirror symmetry keeps two corners intact. Thus, the bulk-boundary correspondence relates the mirror-graded bulk invariant to topologically protected bound states at these corners.

Since the diagonal mirror symmetries \emph{commute} with the chiral symmetry, $H(\vex k'_{mj}), M'_j$ and $C$ can all be block-diagonalized simultaneously. In each block $C_{j\pm}$ are chiral symmetries of $H_{j\pm}(\vex k'_{mj})$, for which one may define, in the usual way, the winding number $W$ of a chirally symmetric Hamiltonian. In the chiral basis, such a Hamiltonian $H(\zeta)$ parametrized by a compact variable $\zeta$, is off-diagonal,
\begin{equation}
H = \begin{pmatrix} 0 & h^\dagger \\ h & 0 \end{pmatrix} \Rightarrow
W[H] = \frac1{2\pi i} \oint \frac{\partial  \log \det h(\zeta)}{\partial \zeta} d\zeta.
\end{equation}
Therefore, 
\begin{equation}
\nu_{j\pm} \equiv W[H_{j\pm}(\vex k'_{mj})].
\end{equation}
We note that the presence of two anticommuting diagonal mirror symmetries dictates $\nu_{j+} = -\nu_{j-}$, thus $\nu_j = \nu_{j+}$. 

Indeed, in Ref~\onlinecite{TopoMatter}. it was shown that this mirror-graded invariant captures the higher-order topology of the topological quadrupole phase. In Section~\ref{sec:HOFT}, we shall use this characterization to demonstrate the higher-order Floquet topological phases. 

\vspace{-3mm}
\subsection{Floquet Theory of Periodically Driven Model}
Here, we review Floquet theory and fix our notation to describe the periodic dynamics of the system. The driven model has a periodic Hamiltonian $H(t)$, with period $\tau=2\pi/\Omega$, via a time-periodic hopping modulation $f_\vex r(t)$. The dynamics is given by the time-ordered evolution operator
\begin{equation}
U(t',t)= \Texp\left[-i\int_t^{t'} H(s) ds\right].
\end{equation}
We shall use Floquet theory to separate the motion within a drive cycle and the stroboscopic evolution of successive cycles.

According to Floquet theorem~\cite{Flo83a}, the solutions of the Schr\"odinger equation take the form $e^{-i\epsilon_\alpha t}\ket{u_\alpha(t)}$, where quasienergies $\epsilon_\alpha\in[-\Omega/2,\Omega/2]$ are conserved and the periodic Floquet modes $\ket{u_\alpha(t)} = \ket{u_\alpha(t+\tau)}$ are eigenstates of the Floquet  evolution operator,
$
U_F(t) \ket{u_\alpha(t)} = e^{-i \tau \epsilon_\alpha} \ket{u_\alpha(t)}.
$
Here,
\begin{equation} 
U_F(t) \equiv U(t+\tau,t) =: e^{-i\tau H_F(t)},
\end{equation}
defines the Floquet Hamiltonian $H_F(t)=H_F(t+\tau)$. 

\vspace{-3mm}
\subsection{Floquet Topological Invariants}
Consider an instantaneous Hamiltonian, which has a unitary temporal mirror symmetry,
\begin{equation}
M_t^\nodag H(t_m+t) M_t^{-1} = H(t_m-t),
\end{equation}
around reflection-symmetric times $t_m = 0$ and~$\tau/2$. Under this symmetry, the Floquet operator is mapped to $\tilde U^\dagger_F(t_m) = M_t U_F(t_m) M_t^{-1} $, where $\tilde U_F$ is obtained from $\tilde H = - H$. In the presence of a chiral symmetry, $\tilde H = C^{-1}HC$, and $\tilde U^\dagger_F = C U^\dagger_F C$. Thus, the Floquet Hamiltonian at reflection-symmetric times is chirally symmetric,
\begin{equation}
C_t^\nodag H_F(t_m) C_t^{-1} = - H_F(t_m)
\end{equation}
with $C_t = M_tC$, if $[C,M_t]=0$, and $C_t = i M_t C$ if $\{C,M_t\} = 0$. Moreover, any symmetry $S$ of the instantaneous Hamiltonian, $[S,H(t)]=0$, is also a symmetry of the Floquet Hamiltonian, $[S,H_F(t)]=0$, at any initial time. In the following, we will assume that the temporal and spatial mirror symmetries commutes, $[M_t,M_j] = 0$. Then, $C_t$ will have the same commutation algebra as $C$ with the spatial mirror symmetries.

Consistent with these symmetries, the Floquet Hamiltonians $H(t_m)$ have their own stable topological indices, $\nu_F(t_m)$. Accordingly, \emph{two} topological invariants are defined~\cite{asboth2014} 
\begin{equation}\label{eq:Ftop}
\nu_{F}^\eta = \frac{\nu_{F}(0) + \eta \nu_{F}(\tau/2)}2,
\end{equation}
associated with quasienergies $\epsilon^\eta = 0, \Omega/2$, where the sign $\eta = e^{-i\tau \epsilon^\eta } = \pm1$.
For example, a chirally symmetric Floquet Hamiltonian has a topological invariant defined as its winding number $\nu_{F}(t_m) = W[H_F(t_m)]$. Indeed, any other topological invariant of a system with chiral and time-mirror reflection symmetry, such as the invariants defined through nested Wilson loops or mirror-graded eigenspaces, can be converted in this fashion to Floquet topological invariants characterizing the topology of the periodic dynamics.


We will now use this method to study the higher-order Floquet topologies induced in our driven model.

\vspace{-3mm}
\section{Higher-order Floquet topological insulator}\label{sec:HOFT}
For simplicity of our presentation, we will focus on two-step drive protocols, in which the hopping modulation $f$ periodically switches between two values $f_{t1}$ with a duration $\tau_1$, and $f_{t2}$ with a duration $\tau_2 = \tau-\tau_1$. This is simple enough to allow analytical and exact numerical calculations, yet rich enough to demonstrate the physics of interest. This two-step protocol is time-mirror symmetric with $M_t=\id$ and reflection-symmetric times in the middle of each step, which we set at $0$ and $\tau/2$, respectively. The Floquet evolution operator at these times are
\begin{equation}
U_F(0) = U_{t_1}^\dagger U_{t_2}^{\nodag 2} U^\nodag_{t_1}, \quad U_F(\tau/2) = U_{t_2}^\dagger {U_{t_1}^{\nodag 2}} U\nodag_{t_2},
\end{equation}
where $U_{t_\mathcal{s}} = e^{-i(\tau_{\mathcal{s}}/2) H_{t_\mathcal{s}}}$, with
$
H_{t_\mathcal{s}} = H(\vex w_{t_\mathcal{s}},\vex f_{t_\mathcal{s}}),
$
for drive step $\mathcal{s}=1,2$. The two Floquet evolution operators are related by the half-cycle micromotion operator $\Phi = U_{t_2}U_{t_1}$ as $U_F(0) = \Phi^\dagger U_F(\tau/2) \Phi $.

\begin{figure}[t]
   \centering
   \includegraphics[width=3.4in]{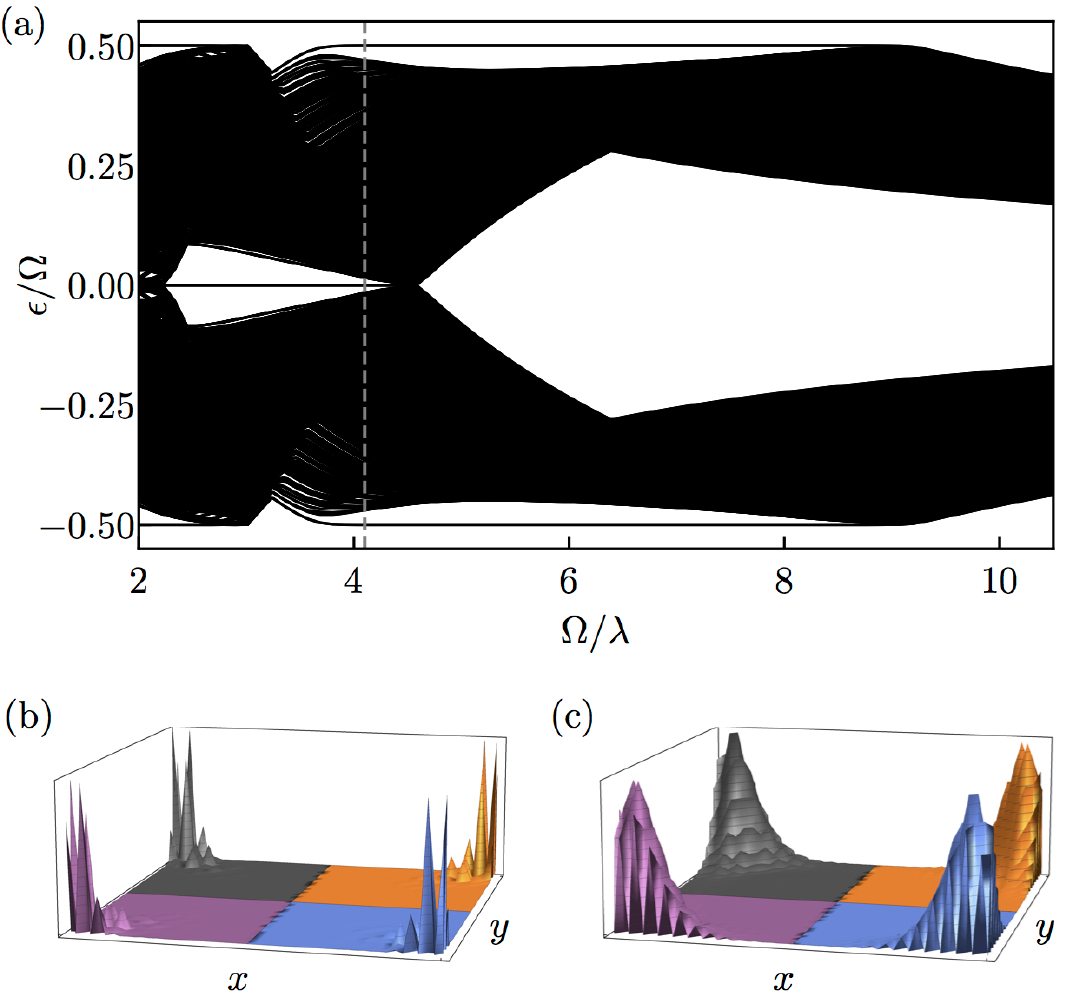} 
   \caption{Floquet spectrum of the driven model. (a) The quasienergies of a system with open boundary conditions vs. drive frequency, $\Omega$. The system has $50\times50$ sites and drive parameters, 
      $w_{1t_\mathcal{s}} = w_{2t_\mathcal{s}}=w_{t_\mathcal{s}}$ and $f_{t_\mathcal{s}} = |f_{t_\mathcal{s}}|e^{i\pi/4}$ for time steps $\mathcal{s}=1,2$, with 
      $w_{t_1}=2.25$, $|f_{t1}| = \sqrt2/1.8$, $w_{t_2} = 1.005$, $|f_{t2}|=\sqrt2/201$, and duration $\tau_1=\tau_2=\pi/\Omega$, are chosen to be in the trivial phase of the instantaneous Hamiltonian and respect diagonal mirror symmetries. The frequency is shown in units of inter-unit-cell hopping $\lambda=w_{\mathcal{s}}(1-f_{\mathcal{s}})$, taken to be the same for both drive steps. Floquet bound states are seen in different ranges of frequency at $\epsilon^+=0$ and $\epsilon^-=\Omega/2$.  The probability density of the four degenerate Floquet bound states, 
      at $\epsilon^+$ (b) and $\epsilon^-$ (c) are shown. The drive frequency here is $\Omega / \lambda = 4.1$, marked with a dashed line in (a).}
   \label{fig:HOFTspec}
\end{figure}

\vspace{-3mm}
\subsection{Floquet corner states}
We first present the numerical evidence for higher-order topological phases in open boundary conditions. In Fig.~\ref{fig:HOFTspec}, we show the Floquet spectrum of the driven model as the frequency of the drive is lowered. The hopping modulations $\vex f_{t1}$ and $\vex f_{t2}$ are both chosen to be in the trivial phase of the static model. At sufficiently high frequencies, the periodic dynamics is described by a time-independent Floquet Hamiltonian $H_F = \overline H$, where $\overline H = H^{(0)} = \int_0^\tau H(t) dt/\tau$ is the average Hamiltonian over a cycle. Thus, the Floquet topology is the same as the equilibrium topology of this average Hamiltonian, which is trivial in our case. 

As the frequency is lowered, the coupling between Floquet modes in different Floquet zones increases and the Floquet topology can change when the quasienergy gap at the Floquet zone edge ($\epsilon^- = \Omega/2$) and/or center ($\epsilon^+=0$) close, either in the bulk or at the edges. With our choice of drive parameters, the changes of topology are accompanied by quasienergy gap closings in the bulk as seen in Fig~\ref{fig:HOFTspec}(a). These nontrivial topologies host Floquet bound states at the corners of the system, see Fig~\ref{fig:HOFTspec}(b), which signal the higher-order nature of the Floquet topological phase. 

At reflection-symmetric times, the Floquet Hamiltonian has all the same symmetries as the instantaneous Hamiltonian. Thus, when $H_{t_\mathcal{s}}$ have diagonal mirror symmetries, so do the Floquet Hamiltonians $H_F(t_m)$. Thus, the topological phase transitions can only occur along paths in parameter space $(\vex w, \vex f)$, which respect these symmetries. With diagonal mirror symmetries, these transitions can only happen when the bulk gap closes.

\vspace{-4mm}
\subsection{Floquet Hamiltonians of Two-Step Drive}
The algebraic form of the Floquet Hamiltonians is dictated by the symmetries to be,
\begin{equation}
H_F(t_m) = A_1 D_{F1}(\vex k) + A_2 D_{F2}(\vex k). 
\end{equation}
The main difference with $H(t)$ here is that the operators $D_{Fj}=|d_{Fj}|e^{i\phi_{Fj}C_j}$ now depend on both components of the lattice momentum $\vex k$. While the algebraic structure of the Floquet Hamiltonian $H_F(t_m)$ is the same as the instantaneous Hamiltonian, its dependence on lattice momentum $\vex k$ and parameters $\vex w$ and $\vex f$ can be quite different and complicated. In this way, the periodic drive generates a whole family of different Hamiltonians consistent with the algebra of symmetries. 

In order to see this explicitly, we calculate the Floquet Hamiltonians for the two-step drive.
For two Hamiltonians, $H_{t_1}$ and $H_{t_2}$, of the two-step drive, we have
$
\{H_{t_1}, H_{t_2} \} = 2 \mathcal{d}_{t_1}\cdot\mathcal{d}_{t_2}. 
$
Using these commutation relations, one may find closed-form expressions of the Floquet Hamiltonians. We present the details in Appendix~\ref{app:HF} and summarize the results here. The quasienergy bands \( \pm \epsilon \), with $\epsilon>0$, are found from 
\begin{align}
\cos(\tau\epsilon) 
	&= \cos(\tau_1E_{t_1})\cos(\tau_2E_{t_2}) \nonumber \\
	&~~~- (\hat{\mathcal{d}}_{t_1} \cdot \hat{\mathcal{d}}_{t_2}) \sin(\tau_1E_{t_1})\sin(\tau_2E_{t_2}),
	\label{eq:eFfull}
\end{align}
The 4-vector for the Floquet Hamiltonian is found to be
\begin{equation}
\mathcal{d}_F(t_m) = \frac{\epsilon}{\sin(\tau\epsilon)} \left[ c_1(t_m) \hat{\mathcal{d}}_{t_1} + c_2(t_m) \hat{\mathcal{d}}_{t_2} \right],
\label{eq:dFfull}
\end{equation}
where the unit vector \( \hat{\mathcal{d}} = \mathcal{d}/E \), and
\begin{align}
c_1(0) 
	&= \sin(E_{t_1}\tau_1)\cos(E_{t_2}\tau_2) \nonumber\\
	&~~~ - (\hat{\mathcal{d}}_{t_1} \cdot \hat{\mathcal{d}}_{t_2}) \left[1- \cos(E_{t_1}\tau_1) \right] \sin(E_{t_2}\tau_2),
	\label{eq:c10} \\
c_2(0)
	&= \sin(\tau_2E_{t_2}).
	\label{eq:c20}
\end{align}
To obtain the expressions for $t_m=\tau/2$, one swaps $1\leftrightarrow 2$ everywhere in Eqs.~(\ref{eq:c10}) and~(\ref{eq:c20}).

\vspace{-3mm}
\subsection{High-Frequency Approximation}
In the high-frequency limit, we can also use the Baker-Campbell-Hausdorff formula,
$$
e^Xe^Y = e^{X + Y + \frac12 [X, Y] + \frac1{12}([X,[X,Y]]+[Y,[Y,X]])+\cdots},
$$
to find 
\begin{equation}\label{eq:high}
H_F(t_m) = \overline{H} + e^{i\Omega t_m} \frac{\tau_1\tau_2}{24} [\overline{H},[H_{t_1}, H_{t_2}]]  + O(\tau^3),
\end{equation}
where $\overline H = (\tau_1 H_{t_1} + \tau_2 H_{t_2})/\tau$ is the average Hamiltonian, and $e^{i\Omega t_m} = \pm1$ for the two reflection-symmetric times. For example, taking $w_x = w_y = w$ to be time-independent, we find $|d_{Fj}|e^{i\phi_{Fj}}$ given by Eq.~(\ref{eq:dj}) with $w$ and $f_j$ replaced with $w_F$ and $f_{Fj}$, where
\begin{equation}\label{eq:HFwF}
\frac{w_F}w = 1 - e^{i\Omega t_m}\frac{w^2\tau_1\tau_2}{12} \sum_{j=1,2} \overline f_j \Delta f_j(1-\cos k_j),
\end{equation}
and
\begin{align}\label{eq:HFfF}
\frac{w_F}w f_{Fj} &= \overline f_j - e^{i\Omega t_m}\frac{w^2\tau_1\tau_2}{12}\bigg[\Delta f_j \sum_{j=1,2} (1+\cos k_j) \nonumber \\
&~~~~~~~~~+ (-1)^j f_{21} (1-\cos k_{\bar j})\bigg].
\end{align}
Here, the average $\overline{f} = (\tau_1 f_{t_1} + \tau_2 f_{t_2})/\tau$, the difference $\Delta f = f_{t_2} - f_{t_1}$, and $f_{21} = \im (f_{t_2}^*f_{t_1}^\nodag)$.
Details of this calculation are also presented in Appendix~\ref{app:HF}.

\begin{figure}[t]
   \centering
   \includegraphics[width=3.4in]{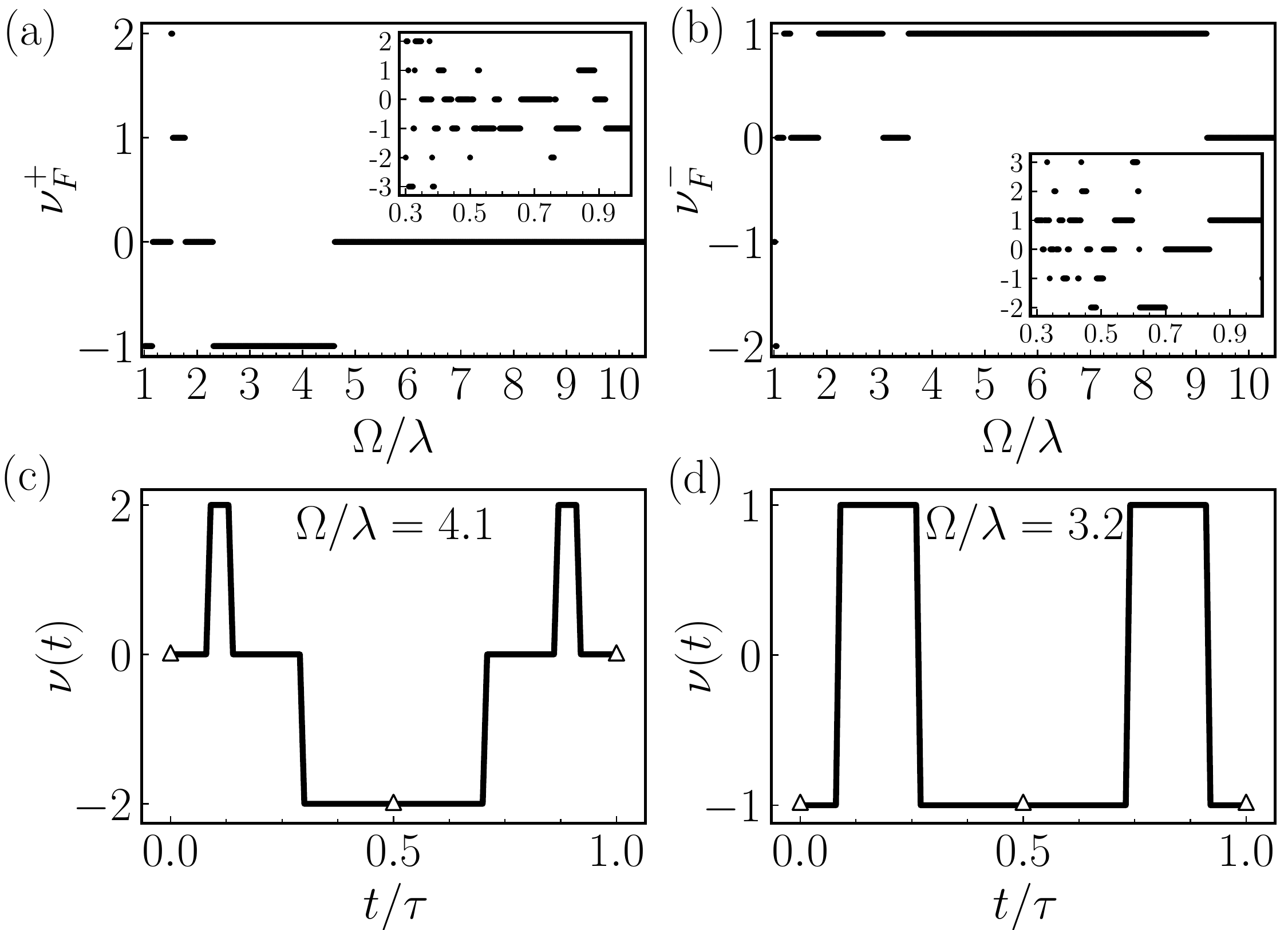} 
   \caption{Floquet topological invariants for the driven model with diagonal mirror symmetries and periodic boundary conditions. The drive parameters are the same as in Fig.~\ref{fig:HOFTspec}. The mirror-graded Floquet topological invariants $\nu_F^\pm = \frac12[\nu_F(0) \pm \nu_F(\tau/2)]$ for $\epsilon^+=0$ (a) and $\epsilon^-=\Omega/2$ (b) vs. drive frequency. The insets show a magnification in a lower frequency range. (c,d) The variation of mirror-graded topological invariant of the Floquet Hamiltonian, $\nu_F(t)$, as a function of the initial time in the cycle, $t$ (see text for definitions). The stable mirror-graded invariants, $\nu_F(t_m)$, at reflection-symmetric times $t_m=0$ and $\tau/2$ are marked with triangles.}
   \label{fig:HOFTwind}
\end{figure}

\vspace{-3mm}
\subsection{Mirror-Graded Floquet Topological Invariants}
Choosing the instantaneous Hamiltonian to have diagonal mirror symmetries, i.e. $w_1=w_2$ and $f_1 = f_2$ for both steps of the drive, we obtain Floquet Hamiltonians at reflection symmetric times, which also have the diagonal mirror symmetries. 
Thus, we can define stable mirror-graded topological invariants for these Floquet Hamiltonians.
Following the definition of Eq.~(\ref{eq:Ftop}), we can thus compute the mirror-graded Floquet topological invariants of the periodic drive.

In Fig.~\ref{fig:HOFTwind}, we plot the mirror-graded Floquet topological invariants of the model with periodic boundary conditions and the same parameters as in Fig.~\ref{fig:HOFTspec}. In Fig.~\ref{fig:HOFTwind}(a,b), the Floquet invariants are shown as a function of frequency. (The numerical calculation and the one using the analytical expression of Floquet Hamiltonians agree precisely.) They correctly show topological phase transitions when bulk gap closes. They also correctly predict the presence of Floquet corner states in both quasienergy gaps around $\epsilon^+=0$ and $\epsilon^-=\Omega/2$ for the system with open boundary conditions.

In order to visualize how mirror-graded Floquet topological invariants arise in the periodic dynamics, we show in Fig.~\ref{fig:HOFTwind}(c,d) for two representative cases, the evolution of the winding number $\nu(t)$ of $H_F(t)$ as the initial time is varied through a cycle. This winding number is calculated as $\nu(t) =\frac12[ \nu_+(t) - \nu_-(t)]$, with
$$
\nu_\pm(t) = \frac1{2\pi i} \oint \frac{\partial}{\partial k}   \log \det [h_\pm(k,t)] dk,
$$
where $h_\pm(t)$ are the off-diagonal elements of $H_{F}(t)$ along $\vex k'_{m2}=(k,k)$ projected on the eigenspaces of $M'_2$ with eigenvalues $\pm1$. Note that  these projections are chirally symmetric at reflection symmetric times $t_m$ only. Thus, $\nu(t)$ is a stable topological invariant only at $t=t_m$. This is why it changes as $t$ is varied through the cycle. However, by plotting $\nu(t)$ for all times in the cycle, we can track its changes more easily and obtain the mirror-graded Floquet topological invariants with confidence.
%
\begin{figure}[b]
   \centering
   \includegraphics[width=3.465in]{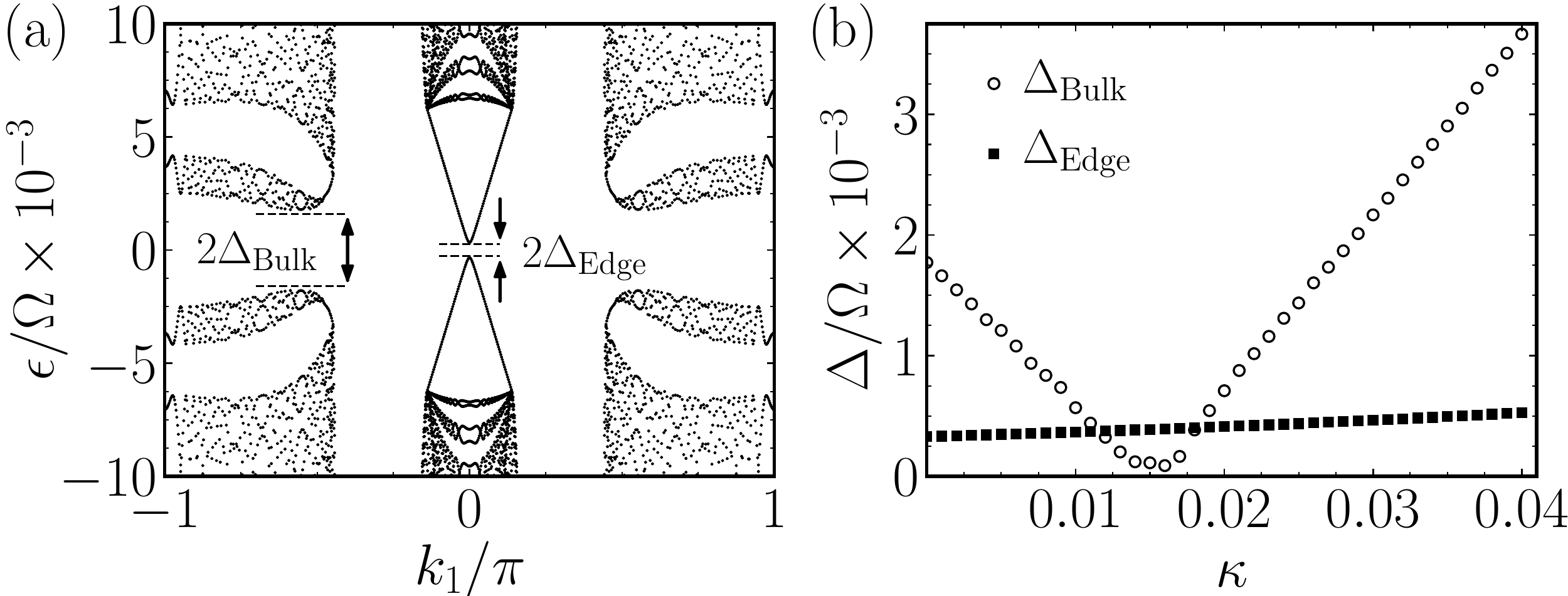} 
   \caption{\noteI{(a) Quasienergy spectrum defined on a strip geometry as a function of momentum $k_1$ parallel to the strip edge for diagonally symmetric model. The drive parameters are the same as in Fig.~\ref{fig:HOFTspec} for $\Omega/\lambda = 1.52$. The edge and bulk gaps are indicated by the arrows. (b) Edge and bulk gap vs. diagonal symmetry breaking strength for $\vex f_\mathcal{s} = |f_{t_\mathcal{s}}| (1+\kappa, 1) /\sqrt{2} $.}}
   \label{fig:strip_geometry}
\end{figure}
%
In particular, in certain ranges of frequency, we find $\nu_F^\pm=\pm1$. This  corresponds to two Floquet corner states at each corner for the open system as shown in Fig.~\ref{fig:HOFTspec}(b). We call this, in accord with previous literature, a higher-order \emph{anomalous} Floquet topological phase. 

As the frequency is lowered further, we find a whole zoo of integer winding numbers, see the insets of Fig.~\ref{fig:HOFTwind}(a,b). These mirror-graded Floquet topological invariants show fluctuations similar to those found previously for one-dimensional chirally symmetric driven models~\cite{vega2018}.

\subsection{\noteI{$\mathbb{Z}$ Invariants with Diagonal Symmetry Breaking}}\label{sec:Zdmb}
\noteI{
Remarkably, we also find winding numbers $|\nu_F^\pm|>1$. For instance, for drive parameters in Fig.~\ref{fig:HOFTwind} in range of frequencies around $\Omega/\lambda=1.52$ ($\lambda$ is the inter-unit-cell hopping amplitude), we have $\nu^+_F=2$. In this range, the bulk quasienergy spectrum is gapped around $\epsilon^+=0$. Correspondingly, for a system with $120\times120$ sites with fully open boundary conditions, we find two degenerate Floquet bound states localized at \emph{each} corner.}

\noteI{Since mirror-graded invariants are well-defined only when the system has four-fold rotational symmetry in addition to the principal mirror symmetries, it is natural to ask whether these multiple corner states survive in the absence of diagonal mirror symmetry. In order to answer this question, we first note that for the system with diagonal mirror symmetries, the bound states on diagonally opposed corners in the direction of $\vex k'_{mj}$ are eigenstates of $M'_j$. Furthermore, at each such corner, $\urcorner_j$, the corner states have the \emph{the same} eigenvalue $\mathcal{m}_{\urcorner_j}$ of $M'_j$. These corner states are mapped by $M'_{\bar{j}}$ to those at $\llcorner_j$, the diagonally opposite corner along $\vex k'_{mj}$; since $\{ M'_j,M'_{\bar{j}},\} = 0$, the corresponding eigenvalues $\mathcal{m}_{\llcorner_j} =  - \mathcal{m}_{\urcorner_j}$. (The details of the arguments proving these statements are presented in Appendix~\ref{app:dmb}.) Now, diagonal mirror symmetry breaking introduces tunneling between corner states with \emph{opposite} eigenvalues of $M'_j$. Therefore, it can only split corner states at different corners. We conclude that the corner states can hybridize and delocalize only if a bulk or edge gap is closed by diagonal symmetry breaking.
}

\noteI{Consider adding diagonal symmetry breaking terms to the Hamiltonian while preserving the principal-axes mirror symmetries, parametrized by the dimensionless symmetry breaking strength $\kappa \geq 0$. Both the time-dependent Hamiltonian and the Floquet Hamiltonian are analytic in $\kappa$ as long as the gaps at $\epsilon^\pm$ remain finite. Since corner states at $\kappa=0$ reside in a finite gap, we expect there to be a finite range of $\kappa>0$ where the gap remains open and, therefore, the number of corner states remains the same as that at $\kappa=0$, where it is given by bulk mirror-graded invariant.}

\noteI{We can also test our expectation by exact numerical diagonalization of the Floquet Hamiltonian in a strip geometry, which gives access to both bulk and edge gaps. In Fig. \ref{fig:strip_geometry} we show our results for the set of parameters giving a bulk invariant $\nu_F^+=2$. As shown, the edge gap remains open (an indeed increases), while the \emph{bulk} gap closes at a critical value of $\kappa$. Thus, the topological phase below this critical value is adiabatically connected to $\kappa=0$ and is described by the bulk mirror-graded invariants for the model with diagonal mirror symmetry.}

\noteII{In Appendix~\ref{app:doubelZ}, we also provide an example of a static \emph{doubled} model without diagonal mirror symmetries that supports more than one zero-energy bound state at each corner. This proves conclusively the inadequacy of a $\mathbb{Z}_2$ invariant to characterize the higher-order topological phases in our model irrespective of the diagonal mirror symmetries.}


\begin{figure}[t]
   \centering
   \includegraphics[width=3.1in]{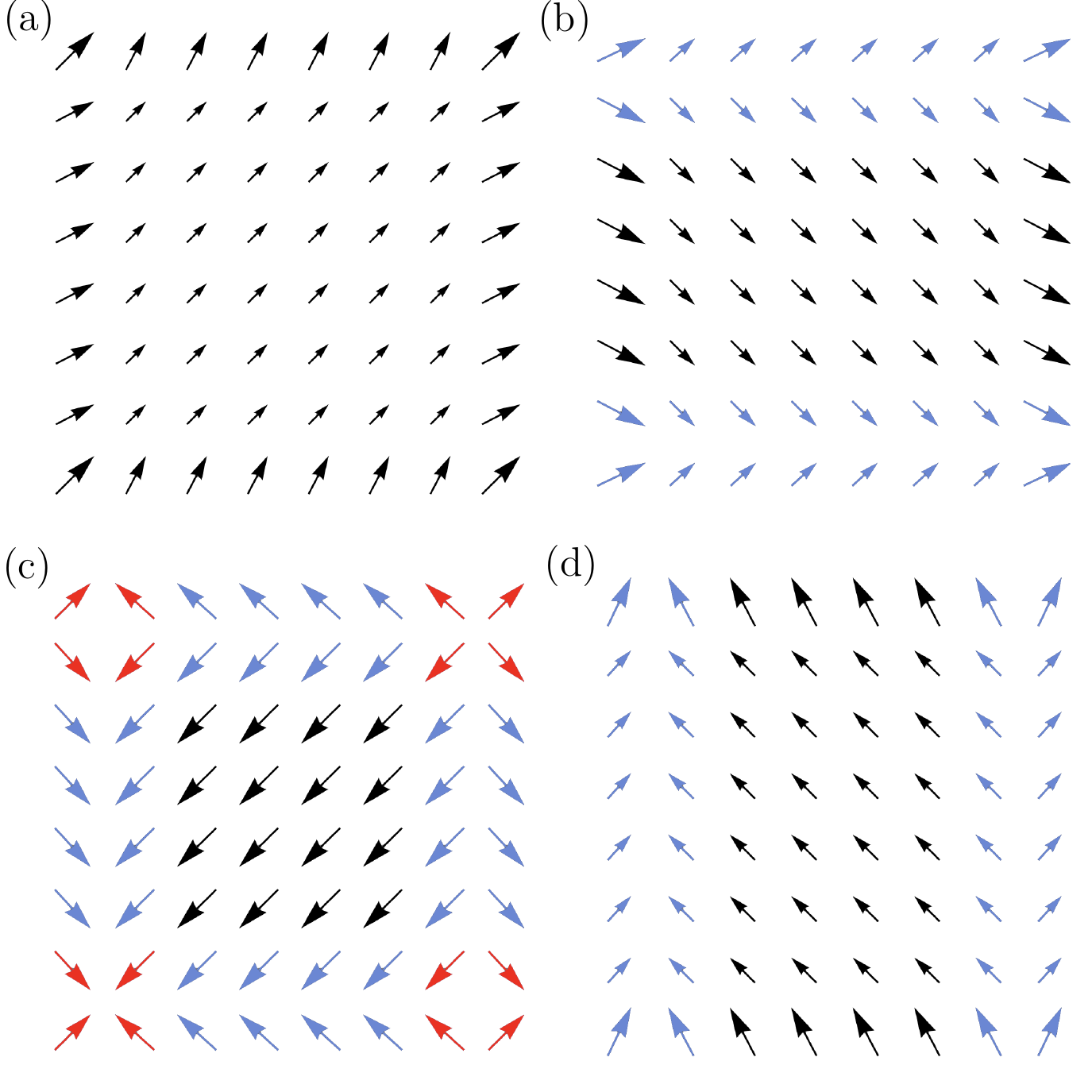} 
   \caption{Defect order parameter, $m(\vex r)$ defined in Eq.~(\ref{eq:defOP}), in the four phases of the static model with open boundary conditions. The trivial phase (a) shows a smooth pattern, while the edge-polarized phases (b, d) show domain walls along the polarized edge. The higher-order topological phase (c), on the other hand, shows domain walls along the edges and vortex defects at all four corners. The domain walls and vortices are colored blue and red for clarity.}
   \label{fig:defOP}
\end{figure}

\begin{figure*}[t]
   \centering
   \includegraphics[width=6.5in]{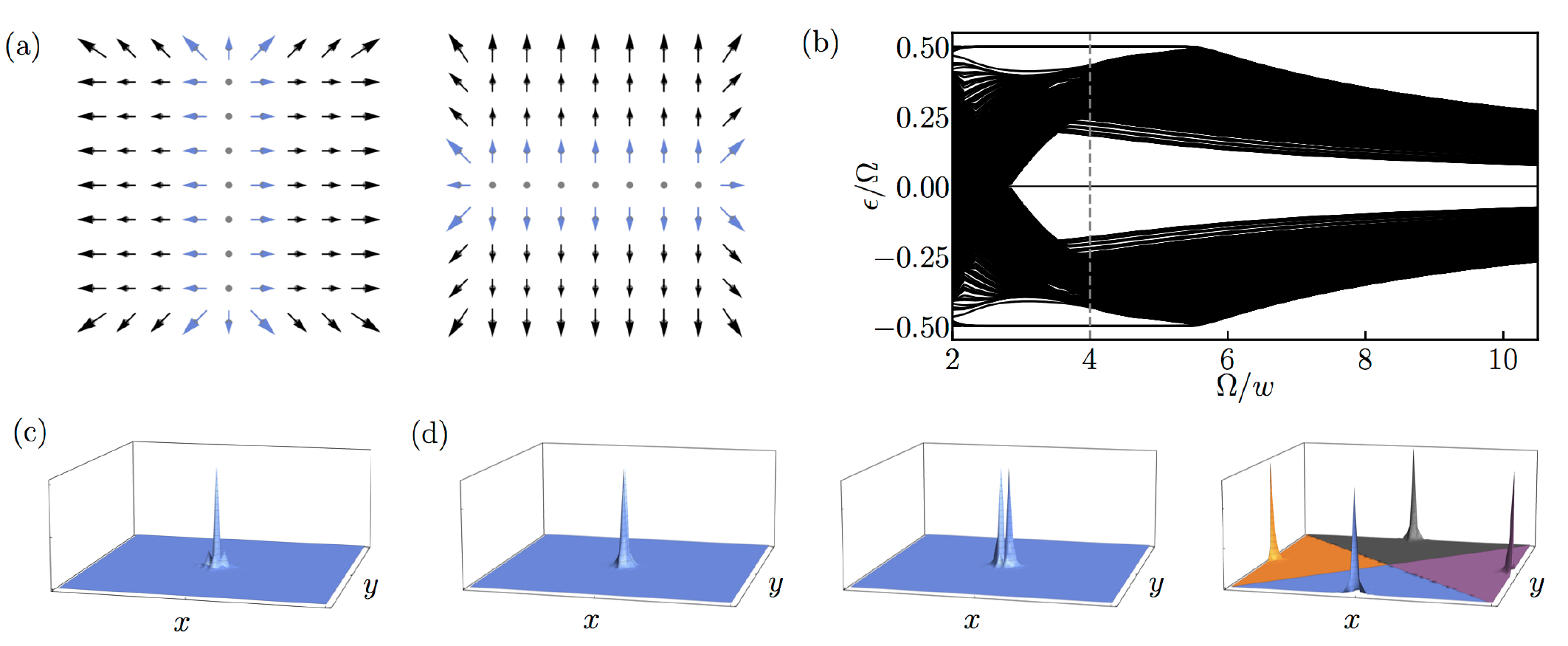} 
   \caption{Stirring protocols and Floquet topological defects. A two-step drive (a) switches between two domain wall configurations with the defect order parameter as shown. The quasienergy spectrum vs. drive frequency for an open system (b) shows a bound state at $\epsilon^+=0$ at arbitrarily high frequencies and, following a bulk gap closing at a lower frequency, a set of 6 degenerate bound states at $\epsilon^-=\Omega/2$. The system size here is $63\times63$ sites and the hopping parameters are $w_1=w_2=w$ and $|f|/w=0.8$. The probability density of the bound states are shown in (c,d). The high-frequency bound state at $\epsilon^+=0$ (c) and two of the lower-frequency ones at $\epsilon^-=\Omega/2$ (d, first two panels) are localized at the intersection of the two domain wall. The other 4 lower-frequency bound states at $\epsilon^-=\Omega/2$ (d, last panel, shown together with different colors) are localized at the intersection of domain walls and the edges of the system.}
   \label{fig:stir}
\end{figure*}

\vspace{-4mm}
\section{Floquet Topological Bulk Defects}\label{sec:stir}
\vspace{-3mm}
\subsection{Vortices in the Static Model}
In previous studies~\cite{seradjeh2008,ChaHouJac08a}, the equilibrium $\pi$-flux dimerized square lattice model was shown to host topologically protected bulk bound states with support at the core of vortex defects in the hopping modulation $f_{\vex r}$. 
Given a pattern of hopping modulations, we may define a defect order parameter~\cite{seradjeh2008},
\begin{equation}\label{eq:defOP}
m(\vex r) = -\sum_\mu \frac1{w_\mu} \eta_{\vex r\mu} w_{\vex r \vex r+\vex e_\mu},
\end{equation}
where $w_{\vex r \vex r+\vex e_\mu}$ and $\eta_{\vex r\mu}$ are the same as in Eq.~(\ref{eq:hop}). The low-energy theory of excitations at half-filling is given by a Dirac Hamiltonian on the background of this defect order parameter~\cite{SerFra08a,RyuMudHou09a},
\begin{equation}\label{eq:lowE}
\mathcal{H} = \sum_{j=1,2} (p_j A_j + m_j B_j),
\end{equation}
where $B_j = i A_j C_j$, we have identified $m= m_1+im_2$ again, and $p_j = -i \partial_j$ is the momentum operator of the excitations. 

Due to this Dirac form, vortex configurations of $m(\vex r)$ bind localized excitations. 
A vortex configuration is realized for $f_{\vex r} = |f(\vex r)| e^{in\arg\vex r}$, where $n\in\mathbb{Z}$ is the quantized vorticity. It supports $n$ mid-gap bound states at zero energy, which are protected by the chiral symmetry $C$ and whose number is a topological invariant related to the index of the Dirac Hamiltonian~\cite{JacRos81a,Wei81a}. The presence of bound states endows a vortex with fractional quantum numbers.

There is indeed a close relationship between these bulk vortices and corner states of the static model. In the bulk, $m(\vex r) = f_{\vex r}$. However, for an open system, hopping amplitudes in the outward directions are set to zero. So, even when $f_{\vex r}$ is uniform in the bulk, the defect order parameter may be nontrivial at the edges and corners. 
Fig.~\ref{fig:defOP} shows plots of $m(\vex r)$ for the case of uniform $f_{\vex r}=f$ with open boundary conditions. Indeed, all phases of the system and, in particular the corner states of the higher-order topological phase, correspond directly to the domain-wall and vortex defects of $m(\vex r)$.

A vortex defect in $m(\vex r)$ need not have full rotational symmetry to support bound states at its core~\cite{seradjeh2008}. For example, a $\mathbb{Z}_4$ vortex defect formed at the intersection of four domains with $f=|f|e^{i\chi}$, $\chi=q\pi/2+\chi_0$, where $0 <\chi_0<\pi/2$ and the domain index $q\in\mathbb{Z}_4$, say, in the clockwise direction, also hosts a bound states at the core. 

We now show that a whole family of bulk vortex defects supporting Floquet topological bound states can be realized dynamically in the driven model.

\vspace{-3mm}
\subsection{Stirring Drive Protocols and Bulk Floquet Topological Defects}
The $\mathbb{Z}_4$ vortex can be generated dynamically by a two-step ``stirring'' drive protocol that switches between two domain-wall configurations,
one with a vertical domain wall with $f = \pm|f|$ on each side, and the other with a horizontal domain wall with $f=\pm|f| e^{2i\chi_0}$ on each side. This is shown in Fig.~\ref{fig:stir}(a) for $\chi_0=\pi/4$. \noteI{(For this example, we chose a system with an odd number of sites in each direction to isolate a single domain wall at the center of the system at each step of the driving protocol.)}
 At high frequency, the Floquet Hamiltonian $H_F = \overline H$ to the lowest order, where $\overline H$ is the average Hamiltonian. Keeping $w_1=w_2=w$ fixed in time, we can see easily that the average Hamiltonian will have four domains intersecting at a vortex defect. Thus, at sufficiently high frequency, we expect to see a Floquet topological bound states localized at the intersection of the two domain walls.

As frequency is lowered, this picture is modified as the topology of quasienergy bands is modified by quasienergy gap closings. In Fig.~\ref{fig:stir}(b), we plot the quasienergy of the driven model with the two-step stirring protocol as a function of drive frequency. As expected, at high frequency there is a Floquet bound state at $\epsilon^+=0$, whose wavefunction is localized at the intersection of the two domain walls, see Fig.~\ref{fig:stir}(c). 
At a lower frequency, a quasi energy gap closing is observed at Floquet zone edge, below which a set of six degenerate Floquet bound states appear at $\epsilon^- = \Omega/2$. In Fig.~\ref{fig:stir}(d) we plot the wavefunctions of these Floquet bound states. \emph{Two} are localized at the intersection of the domain walls. The other four are localized at the intersection of domain walls and edges. 

The above structure can be understood as the dynamical generation of two higher-order anomalous Floquet domains and two trivial ones joined along the oscillating domain walls. \noteI{With our choice of odd number of sites in each direction, bound states associated to nontrivial domains appear at the center instead of the corners of the system.} Indeed, this configuration is similar to that of a static model with a defect order parameter obtained by the superposition of two $\mathbb{Z}_4$ vortices. This would make a vortex with double vorticity that binds two states at its core~\cite{seradjeh2008}. However, in the driven case, the low-frequency driving evidently produces a single vortex structure for $\epsilon^+=0$ with a single Floquet bound states, and a double-vortex structure for $\epsilon^-=\Omega/2$ with two Floquet bound states. These bulk Floquet bound states coexist as steady states of the same model at different quasienergies. This is a novel feature of the stirred Floquet bulk vortices that has no counterpart in the equilibrium model.

\vspace{-5mm}
\section{Discussion and Outlook}\label{sec:discussion}
\vspace{-2mm}
We note that the scheme of dynamically generating bulk vortex defects is not limited to $\mathbb{Z}_4$ vortices. For example, a rotationally symmetric vortex defect can be realized in the high-frequency regime by a continuously stirred domain wall at an angle $\theta=\Omega t$ with the hopping modulation $f(t) = \pm |f| e^{i\theta}$ on each side. Similarly, a multi-step drive stirring a domain wall through $N$ steps, each rotating the domain wall by an angle $\pi/N$, would create a vortex structure with $2N$ domain walls. \noteI{Moreover, the connection between corner and hinge states and bulk defects can also be generalized to any model with a vectorial mass generating the bulk gap. This covers, for example, systems with reflection and/or discrete rotational symmetries~\cite{RyuMudHou09a,FanXiaTon18a}.} As the frequency is lowered in these protocols, we would expect a series of transitions with multiple bulk Floquet bound states appearing at the Floquet zone edge and center. 

The method of creating vortex defects can be easily generalized to higher dimensions. In a three-dimensional model with higher-order topological phases, it would offer a practical way of creating monopoles with fractional quantum numbers using a series of pulses at high frequency. Again, at lower frequencies we expect to find an interesting set of intrinsically non-equilibrium bulk Floquet bound states. By combining these pulses, we envision designing additional dynamics for bulk defects in general and, in particular, adiabatic manipulations that can be useful for quantum information processing.  

A problem that is opened by our work for future study is to find a unified classification higher-order topological phases in and out of equilibrium. The original work focused on nested Wilson loops of Wannier bands, whose structures have been found to be adiabatically connected to the surface Hamiltonians and their higher-order boundaries~\cite{FidJacKli11b}. However, since the interpretation of Wilson loops and their Berry phases in terms of bulk and edge polarization leaves us with only a $\mathbb{Z}_2$  invariant, this approach may not yield the general classification. A more recent study found parallels between the topological structure of Wilson loops and those of Floquet operators, including the presence of anomalous bound states in the Wannier bands~\cite{franca2018}. However, this relationship is not in general well understood. A significant step towards the general classification scheme was taken in Refs.~\onlinecite{max2018} and~\onlinecite{trifunovic2019} by utilizing the classification of topological crystalline phases with second-order spatial symmetries~\cite{ShiSat14a}. Nevertheless, this classification has been achieved for models with only a single spatial symmetry. Thus, the model we study in this paper, with two anticommuting mirror symmetries, is not covered by this classification.

\noteI{Another interesting problem is to find an index for the bulk Floquet topological defects that generalizes that of the static defect order parameter~(\ref{eq:defOP})~\cite{JacRos81a,Wei81a,LuSer14a}. At sufficiently high drive frequency, the Floquet Hamiltonian is given by the average Hamiltonian and, therefore, has a low-energy structure similar to that in Eq.~(\ref{eq:lowE}). However, as the frequency is lowered, the Floquet Hamiltonian develops longer-range hopping terms, which can affect both the low-energy structure and the defect order parameter.} 

In this work, we have presented a periodically driven $\pi$-flux dimerized square-lattice model that realizes the dynamical generation of robust corner states as Floquet bound states. In the presence of four-fold symmetry, these corner states can be classified with two bulk Floquet topological invariants obtained from the winding number of Floquet Hamiltonians graded with a mirror symmetry, along the diagonal lines of the Brillouin zone. Since winding numbers are integers, we conjecture this classification to be given by $\mathbb{Z}\times\mathbb{Z}$, even when the four-fold symmetry is broken, as long as the original mirror symmetries about the unit-cell axes are preserved. We find evidence for this classification at lower frequencies as well as when we consider stirring protocols that dynamically generate vortex defects in the hopping modulations with multiple Floquet bound states in the bulk. Thus, our work paves the way to using simple drive protocols that leverage spatiotemporal inhomogeneities to generate higher-order Floquet topologies. 

\emph{Note added.} In the final stage of preparing this paper, we became aware of two concurrent papers~\cite{huang2018, bomantara2018} discussing high-order Floquet topological phases. {We note that our stirring drive protocols to generate higher-order bulk Floquet topological bound states is not discussed in these papers. The stacking construction~\cite{bomantara2018} is different from our model. In particular, it does not admit a low-energy Dirac Hamiltonian at high drive frequency, and only supports weak higher-order topological phases~\cite{benalcazar2017a,benalcazar2017b}}. \noteI{Also, after our work was completed, Ref. ~\cite{peng2018} presented a classification of higher-order Floquet topological phases with time-glide symmetry.}

\vspace{-5mm}
\begin{acknowledgments}
\vspace{-2mm}
This work was supported in part by the US-Israel Binational Science Foundation grants No. 2014345, the NSF CAREER grant DMR-1350663, and the College of Arts and Sciences at Indiana University. M.R.V acknowledges NSF Materials Research Science and Engineering Center Grant No. DMR-1720595. We also thank the hospitality of CIRM (B.S.) and Max-Planck Institute for the Physics of Complex Systems (M.R.V.), where parts of this work were performed. 
\end{acknowledgments}

\appendix
\vspace{-3.5mm}
\section{Floquet Hamiltonians}\label{app:HF}
In this appendix, we provide some of the details for deriving the Floquet Hamiltonian at reflection symmetric times.

First, for a Hamiltonian of the form~(\ref{eq:piS}), with the 4-vector $\mathcal{d} \equiv (|d_1|\cos\phi_1, |d_1|\sin\phi_1, |d_2|\cos\phi_2, |d_2|\sin\phi_2)$, we have $H^2 = |\mathcal{d}|^2$. Thus,
\begin{equation}
e^{-i t H} = \cos(t E) - i \frac{\sin(t E)}E H,
\end{equation}
where $E = |\mathcal{d}|$. In fact, for any two such Hamiltonians, $H$ and $H'$, we have $\{ H, H'\} = 2\mathcal{d}\cdot \mathcal{d}'$. Using these commutation relations it is easy to calculate the Floquet evolution operator
\begin{equation}
U_F(0) = c_0 + \frac{c_1}{E_{t_1}} H_{t_1} + \frac{c_2}{E_{t_2}} H_{t_2},
\end{equation}
with
\begin{align}
c_0
	&= \cos(\tau_1E_{t_1})\cos(\tau_2E_{t_2}) \nonumber\\
	&~~~ - (\hat{\mathcal{d}}_{t_1} \cdot \hat{\mathcal{d}}_{t_2}) \sin(\tau_1E_{t_1})\sin(\tau_2E_{t_2}),\\
c_1
	&= \sin(\tau_1E_{t_1})\cos(\tau_2E_{t_2}) \nonumber\\
	&~~~ - (\hat{\mathcal{d}}_{t_1} \cdot \hat{\mathcal{d}}_{t_2}) \left[1-\cos(\tau_1E_{t_1}) \right]\sin(\tau_2E_{t_2}), \\
c_2
	&= \sin(\tau_2E_{t_2}).
\end{align}
The unit 4-vector $\hat{\mathcal{d}} = \mathcal{d}/|\mathcal{d}| = \mathcal{d}/E.$ 
Then, we readily obtain Eqs.~(\ref{eq:eFfull})
 and~(\ref{eq:dFfull}) by writing $U_F(0) = e^{-i\tau H_F(0)} = \cos(\tau\epsilon) - i \sin(\tau\epsilon) H_F(0)/\epsilon$. The Floquet Hamiltonian $H_F(\tau/2)$ is obtained by swapping $1 \leftrightarrow 2$.
 
\begin{figure}[t]
   \centering
   \includegraphics[width=3.3in]{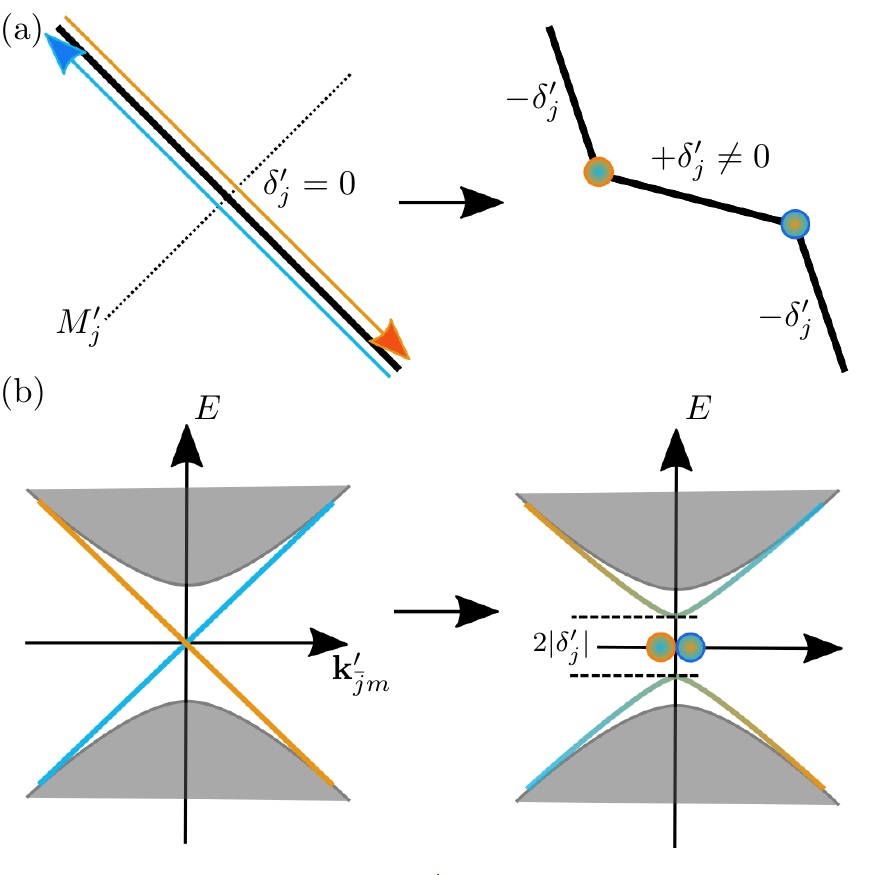} 
   \caption{\noteI{(a) Buckling a diagonally symmetric edge normal to the invariant direction $\vex k'_{mj}$ of the diagonal mirror symmetry $M'_j$ to create two diagonally opposite corners mapped to each other by $M'_{\bar{j}}$. (b) The diagonally symmetric edge has gapless edge modes; the buckled edge has a gapped spectrum away from the corners parametrized by the parameter $\delta'_j$, which changes sign at a corner.
   }}
   \label{fig:dmb}
\end{figure}

Using the commutation relations of matrices $A_j$, and $C_j$ and their products to calculate the high-frequency Floquet Hamiltonian in Eq.~(\ref{eq:high}), we find
\begin{align}\label{eq:4dF}
\mathcal{d}_{F\alpha} = \overline{\mathcal{d}}_\alpha + e^{i\Omega t_m}\frac{\tau_1\tau_2}6\big[ (\overline{\mathcal{d}}{}_\nodag^{\alpha}\hspace{-1mm}\cdot\hspace{-0.5mm}\mathcal{d}_{t_2}^{\alpha})\mathcal{d}_{t_1 \alpha}^\nodag - \{t_1 \leftrightarrow t_2\} \big],
\end{align}
where $\alpha\in\{0,1,2,3\}$ index the components, $\overline{\mathcal{d}}$ denotes the cycle average, and $\mathcal{d}^{\alpha}$ indicates a projection on the three-dimensional space normal to the direction $\alpha$.

Now, we observe that for a given $\vex k$ the Hamiltonian can be cast in a new basis by the gauge transformation $H(\vex k) \to G^\dagger(\vex k) H(\vex k) G(\vex k)$ with 
\begin{equation}
G(\vex k) = e^{-i\sum_{j=1,2}(k_j/4) C_j},
\end{equation}
so that in the new basis $|d_j|e^{i\phi_j} = 2 w_j [\cos(k_j/2) + i f_j \sin(k_j/2)]$. Then, for $w_1=w_2=w$, we have,
\begin{equation}
\mathcal{d} = w\left(\cos\frac{k_1}2,f_1\sin\frac{k_1}2,\cos\frac{k_2}2,f_2\sin\frac{k_2}2\right).
\end{equation}
Replacing $\mathcal{d}_{t_1}$, $\mathcal{d}_{t_2}$, and $\overline{\mathcal{d}}$ in Eq.~(\ref{eq:4dF}) in this form, we obtain Eqs.~(\ref{eq:HFwF}) and~(\ref{eq:HFfF}) in the main text.

\section{Diagonal Mirror Symmetry at Corners}\label{app:dmb}

\noteI{In this Appendix, we will provide rigorous arguments for our statements in Sec.~\ref{sec:Zdmb}, regarding the nature of corner states in a system with diagonal symmetry. First, we consider an edge normal to $\vex k'_{mj}$, the invariant direction of the diagonal mirror symmetry $M'_j$, as shown in Fig.~\ref{fig:dmb}. The momentum $k'_{\bar{j}}$ along the edge is conserved, so the Hamiltonian can be written as $H(x'_j,k'_{\bar{j}})$ where $x'_j$ is the spatial coordinate normal to the edge. Under diagonal mirror symmetries $M'_j H(x'_j,k'_{\bar{j}}) M'_j = H(x'_j,-k'_{\bar{j}})$ and $M'_{\bar{j}} H(x'_j,k'_{\bar{j}}) M'_{\bar{j}} = H(-x'_j,k'_{\bar{j}})$. The edge modes are classified by the mirror-graded topological invariant \noteII{$\nu_{{j}}$}: there are \noteII{$|\nu_{{j}}|$} pairs of gapless edge modes $\psi^\text{edge}_{n\pm}(k'_{\bar{j}})$, \noteII{$n=1,2,\cdots, |\nu_{{j}}|$}. Each pair is an eigenstate of $M'_{\bar{j}}: \psi^\text{edge}_{n\pm}(k'_{\bar{j}}) \mapsto \pm\psi^\text{edge}_{n\pm}(k'_{\bar{j}})$ and is mapped onto itself under $M'_j: \psi^\text{edge}_{n\pm}(k'_{\bar{j}}) \mapsto \psi^\text{edge}_{n\mp}(-k'_{\bar{j}})$. These edge modes are degenerate at $k'_{\bar{j}} = 0$, the invariant momentum under $M'_j$, and the degeneracy is protected by the symmetry of the Hamiltonian under $M'_j$. Projected onto the pair of edge modes in the symmetric direction, we may write $M'_{\bar{j}}=\tau_z$, $M'_{j} = \tau_x$ and the edge Hamiltonian as $H^{\text{edge}}_{j}= k'_{\bar{j}} \tau_z$ with Pauli matrices $\gvex\tau$. 
}

\noteI{Now, consider a small buckling of this edge to form a pair of wide-angle corners as illustrated in Fig.~~\ref{fig:dmb}(a). In this process segments of the edge tilt away from the symmetric direction in opposite ways, resulting in hybridization of the gapless edge modes. This hybridization can be modeled by a small mass term $\delta'_j\tau_y$ in the edge Hamiltonian, whose sign depends on the direction of the tilt. Along the symmetric direction, $\delta_j'=0$. The choice of $\tau_y$ for the mass term is dictated by the fact that under both $M'_j$ and $M'_{\bar j}$, the direction of the tilt is reversed and $\delta'_j \tau_y \mapsto -\delta'_j \tau_y$.
}

\begin{figure}[t]
\begin{center}
\includegraphics[width=3.4in]{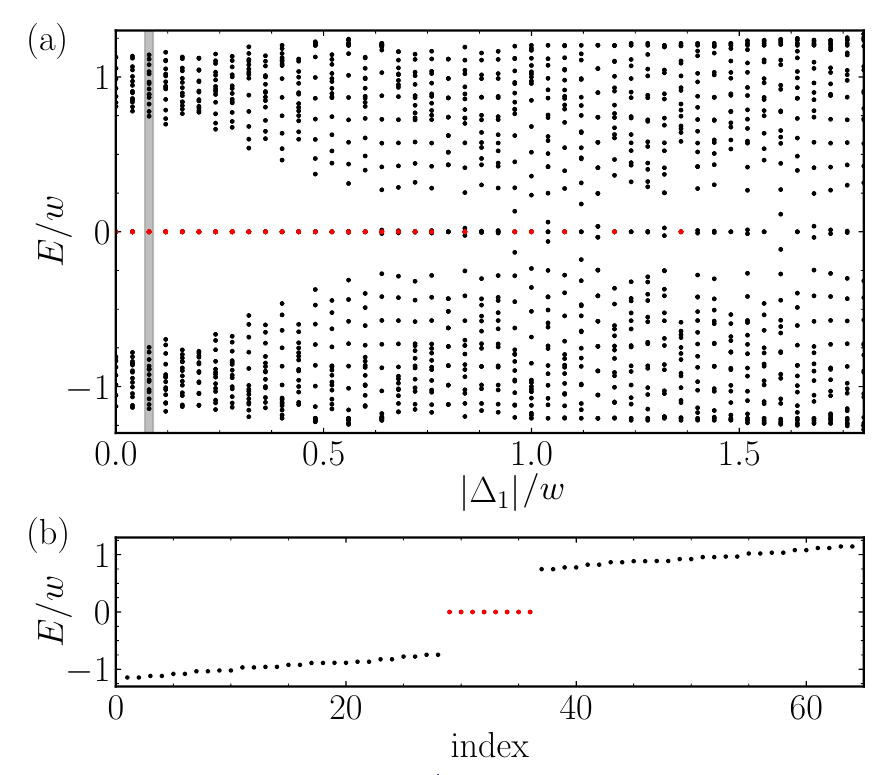}%
\caption{\noteII{The spectrum of the doubled pi-flux dimerized square lattice, Eq.~(\ref{eq:doublepi}), for a system with $50\times50$ sites and open boundary conditions. Sixty-four eigenvalues around zero are shown. The parameters chosen here are $f_1 = -0.4$, $f_2=-0.6$ in each layer, $\Delta_2=0$, and the phase of $\Delta_1$ is chosen at random. Diagonal mirror symmetries are broken everywhere. (a) The bulk and edge gaps remain open over a wide range of $|\Delta_1|>0$ with no mixing between the zero-energy states (red circles). (b) In this range, there are two zero-energy bound states per corner (a total of 8), as shown for $|\Delta_1|/w = 0.08$.}}
\label{fig:doublepi}
\end{center}
\end{figure}

\noteI{Therefore, the buckling with two corners is modeled in the edge Hamiltonian as $H^\text{edge}_j = -i\partial_{y'_j}\tau_z + \delta'_j(y'_j) \tau_y$ with $y'_j$ the spatial coordinate along the edge, such that $\delta'(y')$ switches sign twice at the corners. This results in a pair of \emph{topological kink and anti-kink} in the mass $\delta'_j$ at the corners. Thus, while the edge spectrum is gapped out away from the corners, a pair of zero-energy bound states are localized at the corners with opposite eigenvalues of $M'_j=\tau_x$. This proves our statements in Sec.~\ref{sec:Zdmb} that (i) at a given corner $\urcorner_j$, \emph{all} corner states have the same eigenvalue $\mathcal{m}_{\urcorner_j}$ of $M'_j$, and  that (ii) for the diagonally opposite corner $\llcorner_j = M'_{\bar{j}}$\raisebox{-0.75mm}{$\urcorner_j$}, we have $\mathcal{m}_{\llcorner_j} = -\mathcal{m}_{\urcorner_j}$.
}

\section{\noteII{Inadequacy of the $\mathbb{Z}_2$ Invariant In the Static Model}}\label{app:doubelZ}
\noteII{
In this Appendix we show that the invariant characterizing the higher-order topological phase of the static $\pi$-flux dimerized square lattice is not restricted to $\mathbb{Z}_2$. In order to show this, we show that a doubled model respecting the same algebra of mirror and chiral symmetries can host more than one corner states. When the doubled model also has diagonal mirror symmetries, the number of these corner states is given by the mirror-graded invariant defined in Eq.~(\ref{eq:mirrinv}).
}

\noteII{
The doubled model is defined as
\begin{align}\label{eq:doublepi}
H_2(\vex k) &= H(\mathbf{k}) \otimes \id \nonumber\\
&+  A_1 \otimes(\Delta_{1a}\sigma_x + \Delta_{1b}\sigma_y) \nonumber \\
&+ A_2 \otimes(\Delta_{2a}\sigma_x + \Delta_{2b}\sigma_y),
\end{align}
where $H(\vex k)$ is the Hamiltonian in each layer, Eq.~(\ref{eq:piS}). This can be thought of as two layers of the original model, coupled by two complex tunneling amplitudes $\Delta_1 = \Delta_{1a} + i\Delta_{1b}$ and $\Delta_2 = \Delta_{2a} + i\Delta_{2b}$ connecting the layers in each direction. These tunneling terms are the most general forms that preserve mirror symmetries $M_1\otimes\id$, $M_2\otimes\id$, and chiral symmetry under $C\otimes\id$. Diagonal mirror symmetry is obtained when each layer is diagonally symmetric \emph{and} $\Delta_1 = \Delta_2$.
}

\noteII{
In Fig.~\ref{fig:doublepi}, we show the spectrum of this model on a lattice with open boundary conditions. The lattice has $50\times50$ sites and we have taken the hopping amplitudes in each layer such that they break the diagonal mirror symmetries. Moreover, the diagonal mirror symmetry is also broken in the interlayer tunneling amplitudes as $\Delta_2=0$ throughout. A total of 64 low-lying levels are shown: over a wide range of $|\Delta_1|$, before the bulk gap closes, there are two zero-energy bound states per corner (a total of 8 corner states), proving conclusively that the classification in this model is not $\mathbb{Z}_2$ even when diagonal mirror symmetries are broken. This is consistent with a $\mathbb{Z}$ classification instead.
}


\end{document}